\documentclass[twocolumn,preprintnumbers,showpacs,showkeys,amsmath,amssymb]{revtex4}

\def\ba{\begin{eqnarray}}
\def\ea{\end{eqnarray}}
\def\Eq#1{Eq.\ (\ref{#1})}

\def\>{\rangle}
\def\<{\langle}

\newcommand{\sm}{\mathsf{s}}
\newcommand{\p}{\hat{p}}

\newcommand{\pb}{{\bf p}}

\newcommand{\pbhat}{{\bf \hat{p}}}

\begin{document}

\title{On the Representation of Intermediate States in the Velocity Basis}

\author{B. A. Tay\footnote{Current address: Department of Physics, Faculty of Science,
Universiti Putra Malaysia, 43400 UPM Serdang, Selangor, Malaysia.
Email: \texttt{batay@ph.utexas.edu}} and S.
Wickramasekara\footnote{Current address: Department of Physics,
Grinnell College, Grinnell, Iowa 50112, USA.}}

\affiliation{Center for Particle Physics, The University of Texas at
Austin, Austin, Texas 78712-1081, USA.}

\date{\today}

\begin{abstract}

Unstable state furnishes a semigroup irreducible representation of
the Poincar\'e group. The state vector is represented by a
superposition of energy eigenkets. As a consequence of this
superposition, the state vector can be transformed into the rest
frame through {\it a} Lorentz transformation only when the eigenkets
are labeled by velocity variable, but not momentum variable. We also
clarify the meaning of the velocity variable in the state vector
with respect to the velocity derived from kinematical consideration
of the scattering process.

\end{abstract}

%Use showkeys class option if keyword
\maketitle

%%%%%%%%%%%%%%%%%%%%%%%%%%
%       Sectionn         %
%%%%%%%%%%%%%%%%%%%%%%%%%%
\section{Introduction}

Stable particles furnish the unitary irreducible representation of
the Poincar\'e group \cite{Wigner}. The unitary irreducible
representation is characterized by two Casimir invariants $[m^2,j]$,
i.e. the invariant mass square $m^2$ and spin $j$ of the particles.
Extension of Wigner's idea to encompass unstable particles has been
the subject of many investigations
\cite{Zwanziger,complex1,complex2,complex3,complex4,RGV}. Since
unstable particles decay, probability is not conserved and this
results in a non-unitary irreducible representation for unstable
particles.

Of the many works carried out in this direction, they differ
crucially in the basis employed for the representation space of
unstable particles. Refs. \cite{Zwanziger,RGV} use velocity basis,
whereas Refs. \cite{complex1,complex2,complex3,complex4} resort to
momentum basis for representing the state vector. More specifically,
Refs. \cite{Zwanziger,RGV} define the 4-velocity $\p$ which is
related to the 4-momentum $p$ as \footnote{We do not consider
massless particles.}
%%%
\begin{align}   \label{mc}
        p = \sqrt{\sm} \, \p \,,
\end{align}
%%%
where $\sm$ is the invariant energy square (Mandelstam variable) of
the scattering process. The invariant energy square may acquire
complex value, whereas the 4-velocity is required to remain real.
Ref. \cite{RGV} introduces the additional requirement that the
4-velocity be independent of $\sm$ \cite{partial-wave}. We shall
follow Ref. \cite{partial-wave} and refer to the velocity
representable in the form $\p=p/\sqrt{\sm}$ as the minimally complex
representation. The minimally complex representation has led to a
few interesting consequences. It is the our objective to clarify the
implications of these consequences.

Intuitively, since velocity are kinematically equivalent to
momentum, using velocity as variable in the representation space
should not lead to profound change in the formulation. However,
though the representation is non-unitary, the minimally complex
representation are able to keep the velocity $\bf \p$ and spin $j$
real when the variable $\sm$ is analytically continued into the
complex plane \cite{RGV}. This provides a clean analytic
continuation of the scattering amplitude into the complex
$\sm$-plane when we study resonance phenomena. On the other hand,
complex momentum is inevitable when $\sm$ becomes complex, since
momentum in principle cannot be rendered independent of $\sm$. This
leads to many complex representations in the momentum basis
\cite{complex1,complex2,complex3,complex4} which has no obvious
physical interpretation. This is an important reason for employing
velocity basis in the studies of unstable states.

The requirement that unstable states be transformable by a real
Lorentz transformation to a rest system with 4-momentum of the form,
%%%
\begin{align}   \label{ps}
        p = \sqrt{\sm_r}\, (1,{\bf 0}) \,,
\end{align}
%%%
where $\sm_r$ is complex, has motivated Ref. \cite{Zwanziger} to
introduce real velocity to the representation space of unstable
states. We shall show that unstable states in the momentum basis in
general cannot be transformed into a rest frame by {\it a} Lorentz
transformation, even though a rest frame exists in principle. This
shows that Lorentz covariance is maintained for unstable states in
the velocity basis and velocity basis is favorable over momentum
basis in the description of unstable states.

However, we have to pay a price for this simplicity of description
for unstable states in the velocity basis. By setting the velocity
to be independent of the invariant mass square of the scattering
process, the unstable states are now endowed with a statistical
meaning and devoid of the details of the process (kinematics) from
which they are formed. This amounts to information loss. This fact
will be explained later in the main text. We will also put the
meaning of the velocity variable of the unstable states in proper
perspective.

%%%%%%%%%%%%%%%%%%%%%%%%%%
%       Sectionn         %
%%%%%%%%%%%%%%%%%%%%%%%%%%
\section{Semigroup Representation of Unstable States}

Unstable particles have long been known to be associated with the
resonance poles of the S-matrix \cite{Eden}. These poles are complex
and their positions in the complex plane can be
parameterized\footnote{There has been controversy on the
parameterizations of the decay width and invariant energy square ,
we refer the interested reader to Ref. \cite{Z}.} in terms of the
invariant mass square as $\sm_r = (M-i\Gamma/2)^2 $, where $M$ is
the position of the peak of the resonance with decay width $\Gamma$.

With S-matrix as the starting point, analytically continuing the
scattering amplitude into the complex plane allows the derivation of
the Gamow vector, which is the unstable state associated with the
complex pole of the S-matrix \cite{RGV}. Gamow vector furnishes a
semigroup irreducible representation of the Poincar\'e group and is
characterized by 2 numbers, $[\sm_r, j]$, the complex invariant mass
square $\sm_r$ and the spin of the unstable particle or the partial
wave in which the unstable particle is formed, $j$. In the frame
work of Rigged Hilbert Space \cite{RHS}, the Gamow vector is
represented by the following vector \footnote{The integration range
includes the unphysical negative values of $\sm$ in the lower half
plane of the second sheet. This range is important to ensure an
exact exponential decay of the Gamow vector.}\cite{RGV}
%%%
\begin{align}   \label{GV}
    |{\bf \p} j_3 [\sm_r j]^-\> = \frac{i}{2\pi} \int_{-\infty_{II}}^{\infty} d\sm
    \frac{ |{\bf \p} j_3 [\sm j]^-\> }{\sm - \sm_r} \,,
\end{align}
%%%
where $\pbhat$ is the space component of the 4-velocity and $j_3$ is
the third component of the spin. The minus superscript indicates
that the Gamow vector is analytically continued into the lower half
plane and its time evolution is restricted to the forward light
cone. More properties of the Gamow vector can be found in Refs.
\cite{RGV}.

Since the following discussion on the Lorentz transformation
property of the unstable states is general and need not be
restricted to the Gamow vectors, we shall consider the
phenomenological generalization of \Eq{GV} to intermediate states in
a scattering process,\footnote{We do not consider bound states.} cf.
\Eq{intsket1} and (\ref{intsket2}) below. Intermediate states are
transient states in scattering processes, which may or may not
correspond to resonant states\footnote{For example, the energy may
be well below the threshold to form the unstable particle.}. For
instance, a state $C$ may form temporarily when particles $a$ and
$b$ scatter into particles $d$ and $e$,  i.e.
%%%
\begin{align}   \label{abZcd}
       a+b \rightarrow C \rightarrow d+e\,.
\end{align}
%%%
In a decay chain with 3-body final states, an intermediate states
$D$ may be formed as follows,
%%%
\begin{align}   \label{ZY}
        a+ b  \rightarrow c + D \rightarrow c + e + f \,.
\end{align}
%%%

Let us consider the superposition of energy eigenkets weighted by a
function $w(\sm)$ in the velocity basis,
%%%
\begin{align}   \label{intsket1}
     \int d\sm \, w(\sm)\,| {\pbhat} j_3 [\sm   j]^-\> \,.
\end{align}
%%%
The function $w(\sm)$, for example, may be proportional to the
scattering amplitude. For the Gamow vector, $w(\sm)\sim
1/(\sm-\sm_r)$, which is proportional to the relativistic
Breit-Wigner amplitude. \Eq{intsket1} integrates over all
appropriate values of invariant energy square $\sm$ according to
$w(\sm)$. The corresponding superposition of energy eigenkets in the
momentum basis is
%%%
\begin{align}   \label{intsket2}
     \int d\sm \, w'(\sm)\,| {\pb} j_3 [\sm j]^-\>\,,
\end{align}
%%%
which is weighted by the function $w'(\sm)$.

%%%%%%%%%%%%%%%%%%%%%%%%%%
%       Sectionn         %
%%%%%%%%%%%%%%%%%%%%%%%%%%
\section{Lorentz Covariance of Unstable States in the Velocity Basis}

As pointed out in Ref. \cite{Zwanziger}, a desirable property of the
vectors representing unstable states is their ability to transform
to the rest system under a Lorentz transformation. The vector in the
velocity basis \Eq{intsket1} satisfies this requirement. To see
this, we parameterize the standard boost as \cite{Weinberg}
%%%%%%%%%%%%%%%
\ba   \label{boost}
    L^{\mu}_{\,\,\,\nu}(\p)=\left(
                           \begin{array}{cc}
                             \p^{0} & -{\p_{j}} \\
                             \p^{i} & \delta^{i}_{j}-
               \displaystyle{\frac{{\p^{i}}\,{\p_{j}}}
                    {1+{\p^{0}}}} \\
                           \end{array}
                         \right)\,,
           \quad  i,j=1,2,3\,.
\ea
%%%%%%%%%%%%%%%%%%
Under the Lorentz transformation $L^{-1}(\p)$ (a rotation free
boost), the velocity vector \Eq{intsket1} is transformed into its
rest frame \cite{partial-wave,RGV},
%%%
\begin{align}   \label{phat2qhat_nr}
         U[L^{-1}(\p)]  \int d\sm \, w(\sm) |{\pbhat} j_3 [\sm
        j] ^-\>  =   \int d\sm \, w(\sm) | {\bf 0}  {j}_3 [\sm j]^-\>  \,.
\end{align}
%%%
where $U$ is the unitary representation of the Lorentz
transformation. The 4-momentum of the transformed state is given by
\Eq{ps}. Note that the form of \Eq{intsket1} is retained under the
transformation. Hence we say that the velocity vector is Lorentz
covariant.
%Under a more general Lorentz transformation
%$\Lambda$, the velocity eigenket transforms as
%%%%
%\begin{multline}   \label{phat2qhat}
%        U(\Lambda)  \int d\sm \,w(\sm) |{\pbhat} j_3 [\sm j] ^-\> \\
%        =  \int d\sm \, w(\sm) \displaystyle \sum_{j'_3}
%        D^{j}_{j'_3 j_3}\left(R(\Lambda,\p)\right)
%       |\Lambda \pbhat  {j'}_3 [\sm j] ^-\>  \,,
%\end{multline}
%%%%
%up to the Clebsch-Gordan coefficients $D$ of the Wigner rotation $R$
%\cite{Weinberg}.

Complication arises when we consider the Lorentz transformation of
the momentum vector in \Eq{intsket2}. Let us apply the Lorentz
transformation $L^{-1}(\p)$ to the state vector. Even though all the
constituent momentum kets under the $\sm$ integration in
\Eq{intsket2} have identical momentum $\pb$ (the space components of
the 4-momentum vector $p$), but they have different invariant energy
square when $\sm$ changes. By definition $\pbhat = \pb/\sqrt{\sm}$,
hence for each different $\sm_1 \neq \sm_2 \neq ...$, the
corresponding velocities are different too, i.e. $\pbhat_1 \neq
\pbhat_2 \neq ...$. Therefore, transforming a vector with momentum
$\pb$ to its rest frame
%%%%%%%%%%%%%%%
\ba   \label{p2q}
      \int d\sm \,w'(\sm)\,|{\bf p} j_3 [\sm j] ^-\>
            \rightarrow
                    \int d\sm \,w'(\sm)\,|{\bf 0} j_3 [\sm j] ^-\>\,,
\ea
%%%%%%%%%%%%%%%%%%
can only be achieved with a series of different Lorentz
transformations, i.e. $L^{-1}(\p_i)$, where $i=1,2,3,...$, each
acting separately on $|\pb  j_3 [\sm_i j]\>$, where $\pb =
\sqrt{\sm_i} \pbhat_i$. This violates the requirement of having only
a single Lorentz transformation to bring the momentum vector into
its rest frame. Therefore, \Eq{intsket2} is not a desirable
representation vector for intermediate states.

The complication in the transformation of momentum vector originates
from the fact that the Lorentz transformation (\ref{boost}) is a
function of the 4-velocity $\p$, but not the 4-momentum $p$. This
complication does not arise for stable (asymptotic) states, in both
the velocity basis as well as momentum basis. This is because for
stable particle, $\sm=m^2$ is a constant, where $m$ is the rest mass
of the stable particle. It also follows that the notion of a
superposition of the energy eigenkets over the $\sm$ variable does
not arise in the case of stable particles.

The momentum vector \Eq{intsket2} does not transform covariantly in
the momentum variable under a single Lorentz transformation (since
this will result in a superposition over the momentum variable), but
the velocity vector \Eq{intsket1} does. Even though in our
discussion we have used the minimally complex representation for the
velocity variable, but the result is independent of the functional
dependence of the momentum on $\sm$. Hence in the representation
space of unstable or intermediate states, velocity is the preferred
basis over momentum.

%%%%%%%%%%%%%%%%%%%%%%%%%%
%       Sectionn         %
%%%%%%%%%%%%%%%%%%%%%%%%%%
\section{Consistency of the Minimally Complex Representation with the Kinematics}
\label{linear}

When we define the velocity variable through the minimally complex
representation in \Eq{mc}, we have implicitly assumed that the
velocity is independent of the invariant energy square $\sm$, i.e.
velocity is not a function of $\sm$. This choice is mathematically
legitimate since out of the set of 5 variables $\{{\rm
p_0},\pb,\sm\}$ and with the off mass-shell constraints for unstable
states ${\rm p}^2_0 - \pb^2 = \sm$, we choose the set of 4 variables
$\{\pbhat, \sm \}$ to be independent. The time component of the
velocity is determined through the constraint ${\rm \p}^2_0-\pbhat^2
= 1$.

However, when we consider the kinematics of a scattering event,
there seems to be an apparent contradiction with the assumption of
minimally complex representation \cite{Blum00}. Consider the
kinematics of the resonance formation process\footnote{We consider
resonance formation process though in the original version of the
apparent contradiction, resonance production process \Eq{ZY} is
considered \cite{Blum00}. The apparent contradiction in resonance
production process is considered from a different point of view in
\cite{Kaldass}.}  in \Eq{abZcd}. The rest mass of the stable
particles $a,b,d,e$ are labeled by $m_i$, where $i=a,b,d,e$
respectively. Suppose in the rest frame of particle $a$, it is
bombarded by particle $b$ which has velocity ${\bf \p}_b =\gamma{\bf
v}=(\gamma v,0,0)$, where $v=|\bf v|$ and $\gamma$ is the usual
dilation factor $\gamma=1/\sqrt{1- v^2}$. The 4-momentum of the
unstable state $C$, which equals the total 4-momentum of the system,
is given by $ p_C=p_a+p_b =(m_b \gamma+m_a, m_b \gamma v,0,0) $. The
invariant energy square of the system is $\sm=p_C^2=m_a^2+m_b^2+2m_a
m_b\gamma $. The velocity of the intermediate state can then be
brought into the following form,
%%%%%%%%%%%%%%%%%%%%%%%%%%%%%%
\ba \label{phatz}
        |{\bf\p}_C|=\frac{|\pb_C|}{\sqrt{\sm}} = \frac{\lambda^{\frac{1 }{ 2}}(\sm,
            m_a^2,m_b^2)}{2 m_a\sqrt{\sm}}\,,
\ea
%%%%%%%%%%%%%%%%%%%%%%%%%%%%%%
where the coefficient
%%%
\begin{align}   \label{lambda}
        \lambda(a,b,c)=a^2+b^2+c^2-2(ab+bc+ca)
\end{align}
%%%
frequently occurs in kinematical expression
\cite{Macfarlane,partial-wave}. Since we obtain a $\sm$-dependent
velocity $\pbhat_C$, one may question the validity of introducing a
$\sm$-independent velocity $\pbhat$ that is consistent with the
kinematics.

This controversy arises in the carelessness of identifying the
velocity vector \Eq{intsket1} with the unstable state occurred in a
specific scattering event. As is clear from the Gamow vector, its
construction requires information on the scattering amplitude
encoded in $w(\sm)$, which is derived from a series of different
scattering events that probes through various values of $\sm$ for
the response of the scattering amplitude. Thus, the Gamow vector is
a statistical representation (average sum weighted by $w(\sm)$) for
the whole collection of scattering events, not just a specific one.

A statistical meaning is ingrained implicitly in the construction of
the unstable states. Once the unstable states are formed,
information on the details of the kinematics of each event prior to
the formation of the unstable states are lost. However, the overall
conservation of total momentum, total angular momentum and other
conserved quantum numbers of the scattering process are still
inherited by the unstable states \cite{RGV}. It is in this sense
that the Gamow vector is a universal unstable state. Cares must be
taken in interpreting the velocity variable to avoid apparent
contradiction with the kinematics of a scattering event.

%%%%%%%%%%%%%%%%%%%%%%%%%%
%       Sectionn         %
%%%%%%%%%%%%%%%%%%%%%%%%%%
\section{Conclusion}

We show that the semigroup irreducible representation of the
Poincar\'e group for unstable states in the velocity basis can be
transformed into its rest system by a real Lorentz transformation,
while this is not true for the irreducible representation in the
momentum basis. This is the consequent of 2 facts: (1) the Lorentz
transformation is parameterized by velocity, and (2) to properly
describe unstable states, a superposition of energy eigenkets over a
range of invariant energy square is necessary. As a result, unstable
state transforms covariantly under Lorentz transformation only in
the velocity basis. This distinction between velocity and momentum
basis does not exist when asymptotic states are considered, since
the invariant energy square $\sm=m^2$ of stable particles are
constant. This finding advocates using velocity basis in the studies
of unstable particles, or intermediate states in general.

In scattering experiments, the existence of unstable states can be
reviewed only if the experiments probe across an appropriate range
of invariant energy square. Disconnected isolated events cannot
review the underlying unstable state. Likewise, a mathematical
description of unstable states has to involve the superposition of
eigenkets across a range of energy, weighted by a function
proportional to the scattering amplitude. This construct endows the
Gamow vector with a statistical meaning, and its velocity variable
cannot be identified to the velocity obtained from the kinematical
consideration of a single scattering event.

The minimally complex representation has avoided complex momentum
and complex spin representation of unstable particles. This
simplicity in the description of unstable state has caused the loss
of some information. Except the overall conservation of the total
4-momentum, total angular momentum and conserved quantum numbers,
the details of the process from which the unstable states are formed
are lost. This conforms to the notion of unstable states as
indistinguishable particles that forget its history upon formation,
and the formulation is able to put it on equal footing with stable
particles.

\begin{acknowledgments}
We wish to thank Professor A. Bohm and Professor W. Blum for
bringing our attention to this problem. B.A.T. thanks Dr. T. L. Yoon
for his hospitality when visiting the Department of Physics,
University of Science Malaysia, where part of this paper was
written. B.A.T. thanks Dr. Hishamuddin Zainuddin from the Department
of Physics, Universiti Putra Malaysia and acknowledges the support
of the Ministry of Science, Technology and Innovation, Malaysia
(MOSTI), Postdoctoral Research Scheme (STI).
\end{acknowledgments}

\end{document}